\documentclass[twocolumn,epsfig,a4paper,amsmath,amssymb,showpacs,prl,superscriptaddress]{revtex4-1}
\usepackage{multirow}
\usepackage{amssymb}
\usepackage{amsmath}
\usepackage{amsfonts}
\usepackage{latexsym}
\usepackage{graphicx}
\usepackage{caption}
\usepackage{subcaption}
\usepackage{dcolumn}
\usepackage{bm}
\usepackage{ulem}
\usepackage{color}
\usepackage{soul}
\usepackage{physics}
\usepackage{bbold}
\usepackage{gensymb}
\graphicspath{ {./figs/} }

\newcommand{\be}{\begin{equation}}
\newcommand{\ee}{\end{equation}}
\newcommand{\ba}{\begin{eqnarray}}
\newcommand{\ea}{\end{eqnarray}}
\newcommand{\baa}{\begin{eqnarray*}}
\newcommand{\eaa}{\end{eqnarray*}}

\newcommand{\dg}{^{\dagger}}

\newcommand{\bl}{\left (}
\newcommand{\br}{\right )}

\begin{document}

\title{Many Body Localization in a Two Dimensional Itinerant SYK Model}
\author{T.~Tzen Ong}
\affiliation{RIKEN Center for Emergent Matter Science (CEMS), Saitama 351-0198, Japan}
\affiliation{Department of Applied Physics, University of Tokyo, Tokyo 113-8656, Japan}
\date{\today}

\begin{abstract}
We consider a two dimensional itinerant SYK model of spin-less fermions, with a linear dispersion, interacting via random long range all-to-all interactions. In the large-N limit, we find an asymptotic power series solution of the saddle-point equations, which demonstrate that the SYK interactions drive the formation of localized states in the strong coupling regime. By calculating the current-current correlation function, the ground state is shown to be insulating. Finally, we calculate the out-of-time-correlator (OTOC), which has a $t$-linear growth in time and an associated logarithmic growth of entanglement; thereby proving that the system forms a many body localized phase.
\end{abstract}

\maketitle

The phenomena of localization in a random medium, due to quantum interference, was first discussed by Anderson \cite{AndersonPhysRev1958}, and in the absence of interactions, generic disorder is known to localize all states in one and two dimensions (2-D) \cite{AbrahamsAndersonPRL1979, MirlinRMP2008}. The possibility of a many-body localized (MBL) phase in an interacting disordered system was also pointed out in Ref.~\cite{AndersonPhysRev1958}, which has now been demonstrated rigorously by Basko. {\it et. al.} Ref.~\cite{AltshulerAnnPhys2006}. MBL phases have a number of striking characteristics, including zero DC conductivity, emergent integrability \cite{AbaninPRL2013, HusePRB2014, SondhiHusePRB2013, ImbrieJStatPhys2016}, violation of the eigenstate-thermalization hypothesis \cite{DeutschPRA1991, SrednickiPRE1994, OlshaniiNature2008}. Furthermore, the MBL phase displays a logarithmic growth of entanglement entropy \cite{FazioJStatMech2006, PrelovsekPRB2008, MoorePRL2012, AbaninPRL2013, HusePRB2014, ChenAnnPhys2016, ZhaiSciBul2017}, as opposed to non-interacting Anderson insulators, which display a constant entropy \cite{StolzLettMathPhys2016, ZhaiSciBul2017},  and many-body thermal phases that display a ballistic growth of entanglement \cite{HusePRL2013, DaleyPRX2013}. 

The growth of entanglement can be probed using the out-of-time-correlator (OTOC), $F(t) = \langle \hat{W}\dg(t) \hat{V}\dg(0) \hat{W}(t) \hat{V}(0) \rangle$, which was first discussed in the context of superconductivity \cite{LarkinOvchinnikovJETP1969}, and has been shown to be related to the second R\'{e}nyi entropy \cite{ZhaiSciBul2017}. It has received a lot of attention recently in both condensed matter and high energy physics as a diagnostic of entanglement and chaos \cite{KitaevTalk2014, KitaevKITP2015, ShenkerStanfordJHEP2015, MaldacenaStanfordJHEP2016, MaldacenaStanfordPRD2016, YoshidaQiJHEP2016}. For thermal phases with chaos, the OTOC will eventually decay to zero exponentially at a rate $\lambda_P$, which is the quantum analog of the Lyapunov exponent, with the well-known ``Sachdev-Ye-Kitaev" model (SYK model) \cite{KitaevTalk2014, KitaevKITP2015, SachdevYePRL1993, SachdevPRX2015, MaldacenaStanfordPRD2016} exhibiting one of the fastest growth with a $\lambda_P$ that saturates the upper bound of $\tfrac{2 \pi}{\beta}$ \cite{MaldacenaStanfordJHEP2016}. 

The SYK model allows for the exact solution, in the large-N limit, of a strongly interacting quantum many-body system that is chaotic with near conformal invariance \cite{MaldacenaStanfordPRD2016}, which has also been argued to be holographically dual to an AdS$_2$ black hole \cite{SachdevPRX2015}. This has generated great interest in the community, with several generalizations of the SYK model that have been applied towards study of non-Fermi liquid phases, strongly interacting diffusive phases, as well as the quantum butterfly effect\cite{SachdevGeorgesPRB2017, QiGuSciPost2017, StanfordQiJHEP2017, XuLudwigPRB2017, ZhangChenPRL2017, AltmanPRB2017, BalentsPRB2018}.

Interaction and disorder are treated on an equivalent footing in the SYK model, and since numerical studies of the 1-D random-field XXZ model \cite{HusePalPRB2010, MoorePRL2012, AbaninSerbynPRX2015, ZhaiSciBul2017} have shown that the MBL phase occurs for strong disorder, this motivates a study of a two-dimensional itinerant SYK model. The large-N results that we obtain demonstrate that the system does indeeds form an MBL ground state in the strong-coupling regime.

We consider a 2-D model of spin-less itinerant fermions interacting via a random all-to-all SYK interaction, described by the Hamiltonian,
\ba
H & = & \sum_{\vec{k}} \xi_{\vec{k}} c\dg_{\vec{k}} c_{\vec{k}} + \sum_{ijkl} \frac{J_{ijkl}}{2 N^{3/2}} c\dg_{i} c\dg_{j} c_{k} c_{l}
\ea
Here, $\xi_{\vec{k}} = \pm v_F |\vec{k}|$ is a linear dispersion with the chemical potential $\mu$ set to zero, and the electrons at sites $i = 1 \ldots N$ interact via a four-fermion coupling $J_{ijkl}$ with Gaussian distribution of zero mean and variance $J^2$. 

\begin{figure}[bht!]
\begin{center}
\includegraphics[trim=10mm 50mm 0mm 5mm, clip, height=4.5cm]{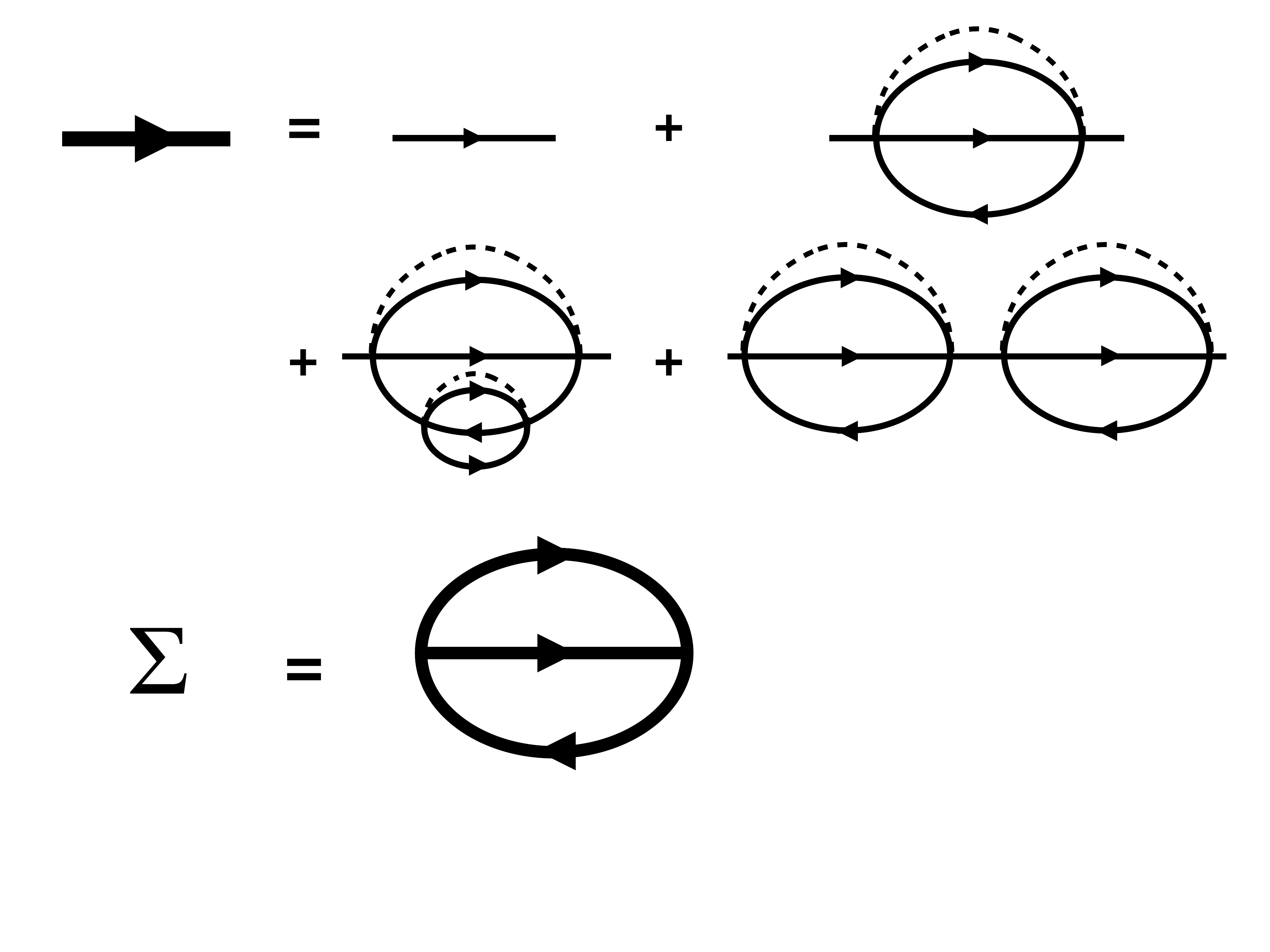}
\end{center}
\captionsetup{justification=RaggedRight, singlelinecheck=false}
\caption{Feynman diagrams for the large-N saddle point effective Green's function, represented by the thick line, which are given by the iterated ``watermelon" diagrams; and the self energy, $\Sigma(\tau)$, which is defined self-consistently in terms of the effective Green's function.}
\label{fig: self energy}
\end{figure}

Applying the standard replica method \cite{Parisi1987} gives $c_{i, a}$, with $a \in [1, n]$ being the replica index, and we then average over the disorder to obtain a replicated action,
\ba
\label{eqn: replica action}
S & = & \int_{0}^{\beta} d \tau \sum_{\vec{k}, a} c\dg_{\vec{k, a}} \bl \frac{\partial}{\partial \tau} + \xi_{|\vec{k}} \br c_{\vec{k}, a} \cr
 & & - \frac{J^2}{4 N^3} \int_{0}^{\beta} d \tau d \tau' \sum_{a b} |\sum_{i} c\dg_{i, a}(\tau) c_{i, b}(\tau')|^4
\ea
Similar to Ref.~\cite{SachdevPRX2015}, the system allows for a replica-symmetric large-N solution, with the following saddle-point equations,
\begin{subequations}
\ba
\label{eqn: saddle point}
G(\vec{k}, i \omega_n) & = & \frac{1}{i \omega_n - \xi_{\vec{k}} - \Sigma(i \omega_n)} \\
\Sigma(\tau) & = & -J^2 G^{(l)}(-\tau) G^{(l)}(\tau)^2
\ea
\end{subequations}
At the large-N saddle point, only the ``watermelon" diagram, as shown in Fig.~\ref{fig: self energy}, contribute to the self-energy after disorder averaging, and similar to Ref.~\cite{KitaevTalk2014, KitaevKITP2015, MaldacenaStanfordPRD2016}, the self-energy is entirely local and depends only upon the local Green's function, $G^{(l)}(\tau) = G(\vec{r} = 0, \tau)$.

Applying a power series solution to $\Sigma(i \omega_n)$, we find the following asymptotic solution (see SOM for details),
\begin{subequations}
\ba
\label{eqn: G local}
\Sigma(i \omega_n) & = & \frac{C_1}{i \omega_n} + \frac{C_2}{\pi} \ln \frac{1 + \frac{D}{C_1} i \omega_n}{1 - \frac{D}{C_1} i \omega_n} \\
\Sigma(\tau) & = & - \frac{C_1}{2} sgn(\tau) + \frac{1}{\pi} \frac{C_2}{\tau} e^{-\frac{C_1}{D} |\tau|} \\
G^{(l)}(i \omega_n) & = & i \frac{C_1}{2 D^2} \frac{1}{i \omega_n} \bl 1 - \frac{i}{\pi} \ln (1 - (\frac{D}{C_1} i \omega_n)^2) \br  \\
 & & + i \frac{C_2}{2 \pi D^2} \ln \frac{(1 + \frac{D}{C_1} i \omega_n)}{ (1 - \frac{D}{C_1} i \omega_n) } \\
G^{(l)}(\tau) & = & -\frac{i}{2 D^2} C_1 sgn(\tau) - \frac{1}{\pi D} \frac{1}{\tau} e^{-\frac{C_1}{D} |\tau|}
\ea 
\end{subequations}
Here, $D$ is the bandwidth, $C_1 = 2 e^{-i \frac{\pi}{4}} \tfrac{D^3}{J}$ and $C_2 = i \tfrac{3}{2} D$. It is clear that these solutions satisfy the saddle-point equations, up to $O(\tfrac{1}{\tau^2} e^{-2 \tfrac{C_1}{D}} |\tau|)$. Hence, $\Sigma(i \omega_n)$ is asymptotically accurate for small $\omega_n \ll D$, i.e. large $\tau$ limit.

The key insight from the solutions is that $G^{(l)}(i \omega_n) \approx i \frac{C_1}{2 D^2} \frac{1}{i \omega_n}$, meaning that the SYK interactions drive the fermions into forming localized bound states in the strong coupling regime, $J \sim D$, with a spectral weight of $\tfrac{D}{J} \sim O(1)$. In addition, the logarithmic term in $\Sigma(i \omega_n)$ indicates the formation of an incoherent background at high energies. We point out that the large-N solution is particle-hole symmetric, as the replica-symmetric action is invariant under charge conjugation, $\mathcal{C}$. 

\begin{figure}[bht!]
\begin{center}
\includegraphics[trim=10mm 70mm 0mm 0mm, clip, width=\columnwidth]{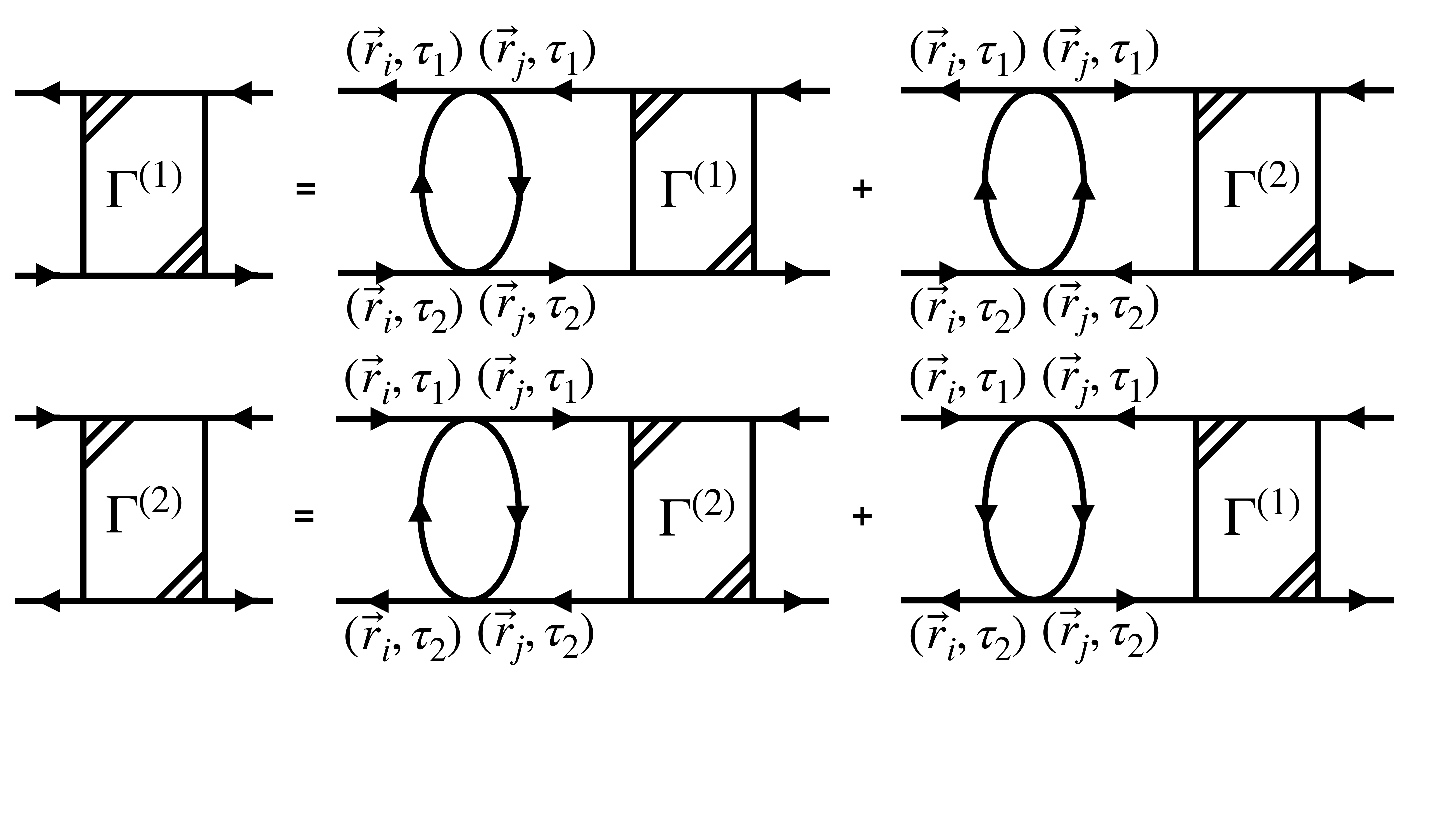}
\end{center}
\captionsetup{justification=RaggedRight, singlelinecheck=false}
\caption{Feynman diagrams for the effective scattering vertices, $\Gamma^{(1)}(\tau_1, \tau_2)$ and $\Gamma^{(2)}(\tau_1, \tau_2)$, which correspond to scattering in the particle-hole and particle-particle channels respectively. Similar to the SYK model, two of the SYK interaction vertices, ($\vec{r}_k$, $\vec{r}_l$), are contracted to form the local bubbles on the rungs, while the external legs,  ($\vec{r}_i$, $\vec{r}_j$) are summed over. The diagrams on the upper and lower left consisting of a particle-hole like bubble, $G^{(l)}(\tau_1 - \tau_2) G^{(l)}(\tau_2 - \tau_1)$,  insertion on the rung, while the upper and lower right diagrams consist of a particle-particle like bubble insertion, $G^{(l)}(\tau_1 - \tau_2)^2$ and $G^{(l)}(\tau_2 - \tau_1)^2$ respectively.}
\label{fig: JJ scattering vertices}
\end{figure}

Since the system forms a lattice of interaction-driven bound states with a high-energy incoherent background, the ground state should be insulating at low energies. This is verified by calculating the longitudinal current-current correlation function, $\chi^{(l)}_{JJ}(\vec{r} - \vec{r}', \tau \tau') = - \langle T_{\tau} j^{x}(\vec{r}, \tau) j^{x}(\vec{r}', \tau') \rangle$, and Fig.~\ref{fig: JJ scattering vertices} shows the ladder diagrams for the scattering vertices.

At O$(1)$, the disorder-averaged SYK interactions give rise to two types of scattering kernels formed by contracting two of the spatial indices to form either a particle-hole like or particle-particle like bubble insertion on the rungs of the ladder, i.e. the left and right diagrams in Fig.~\ref{fig: JJ scattering vertices} respectively. The spatial indices, $\vec{r}_i$ and $\vec{r}_j$, have to be summed over as well at O$(1)$; hence, the scattering vertices only contribute to the uniform $\vec{q} = 0$ component of $\chi^{(l)}_{JJ}(\vec{q}, i \Omega)$.

Since the ground state is particle-hole symmetric, charge conjugation, $\mathcal{C}$, can be applied to the four Green's functions on the legs of the ladder connected to the bubble rung on the upper left diagram for $\Gamma^{(1)}(\tau_1 - \tau_2)$, $\mathcal{C} G(\vec{r} - \vec{r}', \tau - \tau') \mathcal{C}^{-1} = - G(\vec{r}' - \vec{r}, \tau' - \tau)$. This reverses the direction of the Green's functions, thereby giving the lower left diagram for $\Gamma^{(2)}(\tau_1 - \tau_2)$; the upper and lower right diagrams are similarly equivalent. Hence, the scattering kernels of $\Gamma^{(1)}(\tau)$ and $\Gamma^{(2)}(\tau)$ are equivalent under particle-hole symmetry, and therefore, $\Gamma^{(1)}(\tau) = \Gamma^{(2)}(\tau)$.

In Fourier space, the set of coupled integral equations representing the ladder diagrams is given by,
\begin{widetext}
\begin{subequations}
\ba
\Gamma^{(1)}(i \omega_1 - i \omega_2, i \Omega) & = & J^2 \int \frac{d \omega_3}{2 \pi}  2 K^{(1)}(i \omega_1 - i \omega_3, i \Omega) \Gamma^{(1)}(i \omega_3 - i \omega_2, i \Omega) - K^{(2)}(i \omega_1 - i \omega_3, i \Omega) \Gamma^{(2)}(i \omega_3 - i \omega_2, i \Omega) \hspace{1cm} \\
 \Gamma^{(2)}(i \omega_1 - i \omega_2, i \Omega) & = & J^2 \int \frac{d \omega_3}{2 \pi} 2 K^{(1)}(i \omega_1 - i \omega_3, i \Omega) \Gamma^{(2)}(i \omega_3 - i \omega_2, i \Omega)  - K^{(2)}(i \omega_1 - i \omega_3, i \Omega) \Gamma^{(1)}(i \omega_3 - i \omega_2, i \Omega)
\ea
\label{eqn: Vertex integral eqns}
\end{subequations}
\end{widetext}
The scattering kernel $K^{(1)}(i \omega_1 - i \omega_3, i \Omega)$ is given by the $O(1)$ insertion of a particle-hole type bubble rung, $G^{(2l) ph}(i \omega_n)$, which is the Fourier transform of $G^{(2l) ph} (\tau) = G^{(l)}(\tau) G^{(l)}(-\tau)$, along with two legs of the ladder. Similarly, $K^{(2)}(i \omega_1 - i \omega_3, i \Omega)$ involves the insertion of a particle-particle type bubble rung, $G^{(2l) pp}(i \omega_n)$, which is the Fourier transform of $G^{(2l) pp} (\tau) = G^{(l)}(\tau)^2$. Hence, the kernels are,
\begin{subequations}
\label{eqn: curr curr kernels}
\ba
K^{(1)}(i \omega_1 - i \omega_3, i \Omega) & = & H(i \omega_3, i \Omega) G^{(2l) ph}(i \omega_1 - i \omega_3) \hspace{0.5cm} \\
K^{(2)}(i \omega_1 - i \omega_3, i \Omega) & = & H(i \omega_3, i \Omega) G^{(2l) pp}(i \omega_1 - i \omega_3)
\ea
\end{subequations}
where $H(i \omega_n, i \Omega)$ describes the two legs of the ladders connected to the bubble rungs,
\label{eqn: H and G2l}
\ba
H(i \omega_n, i \Omega) & = & \int \frac{d \vec{k}}{(2 \pi)^2} G(\vec{k}, i \omega_n + \frac{i \Omega}{2}) G(\vec{k}, i \omega_n - \frac{i \Omega}{2})  \hspace{0.6cm}
\ea

Since the scattering kernels for $\Gamma^{(1)}(i \omega_n, i \Omega)$ and $\Gamma^{(2)}(i \omega_n, i \Omega)$ are equivalent; hence, $\Gamma^{(1)}(i \omega_n, i \Omega) = \Gamma^{(2)}(i \omega_n, i \Omega) = \Gamma(i \omega_n, i \Omega)$. Eq.~\ref{eqn: Vertex integral eqns} can now be transformed into a differential equation, and up to logarithmic accuracy, it can be solved to give the following solution,
\ba
\Gamma(i \omega, i \Omega) & = & -\frac{\alpha_2}{1 - \frac{\alpha_1}{D^2} \frac{i \omega}{i \Omega} \ln \frac{M(i \omega + i \Omega)}{M(i \omega - i \Omega)} } \\
M(i \omega \pm i \Omega) & = & \bl 1 + \frac{D}{C_1}(i \omega \pm i \Omega)  \br \bl 1 - \frac{D}{C_1}(i \omega \pm i \Omega)  \br \nonumber
\ea
with $\alpha_1 = \tfrac{J^2}{4 \pi} (\tfrac{C_1}{2 D^2})^2$ and $\alpha_2 = \tfrac{J^2}{D}$ (see SOM for details). 

To verify the validity of the solution for $\Gamma(i \omega, i \Omega)$, we check that it satisfies the Ward identity \cite{Abrikosov1975Methods}, which relates the electron self-energy, $\Sigma(i \omega)$ to the scattering vertex, $\Gamma(i \omega)$.
\ba
\label{eqn: Ward}
\frac{d \Sigma(k)}{d i \omega} & = & - \lim_{\Omega \rightarrow 0} \frac{1}{\beta V} \sum_{k_1} G(k_1) G(k_1  + q) \Gamma(i \omega - i \omega_1, i \Omega)  \hspace{0.4cm}
\ea
with $k = (\vec{k}, i \omega)$ and $q = (0, i \Omega)$. The right-hand side of Eq.~\ref{eqn: Ward} is explicitly calculated, and $\lim_{\Omega \rightarrow 0} \frac{1}{\beta V} \sum_{k_1} G(k_1) G(k_1  + q) \Gamma(i \omega - i \omega_1, i \Omega) = \tfrac{C_1}{i \omega^2}$. Therefore, the scattering vertex, $\Gamma({i \omega, i \Omega})$, satisfies the Ward identity, up to logarithmic terms. 

It is now straightforward to evaluate $\chi^{(l)}(\vec{q}=0, i \Omega)$, which consists of the bubble diagram contribution, $\chi^{(l), 0}_{JJ}(\vec{q} = 0, i \Omega)$, as well as the contribution from the vertex correction, $\chi^{(l), v}_{JJ}(\vec{q} = 0, i \Omega)$. To leading order in $\Omega$, we find that the vertex correction cancels the bubble diagram, i.e. $\chi^{(l), v}_{JJ}(\vec{q} = 0, i \Omega) = - \tfrac{e^2 v_F^2}{2} \tfrac{i \Omega}{2 D^2} = - \chi^{(l), 0}_{JJ}(\vec{q}=0, i \Omega)$.  Hence $\chi^{(l)}_{JJ}(\vec{q} = 0, i \Omega) = 0$, for $\Omega \ll D$, as expected for an insulating ground state (see SOM for details).

\begin{figure}[bht!]
\begin{center}
\includegraphics[trim=10mm 70mm 0mm 0mm, clip, width=\columnwidth]{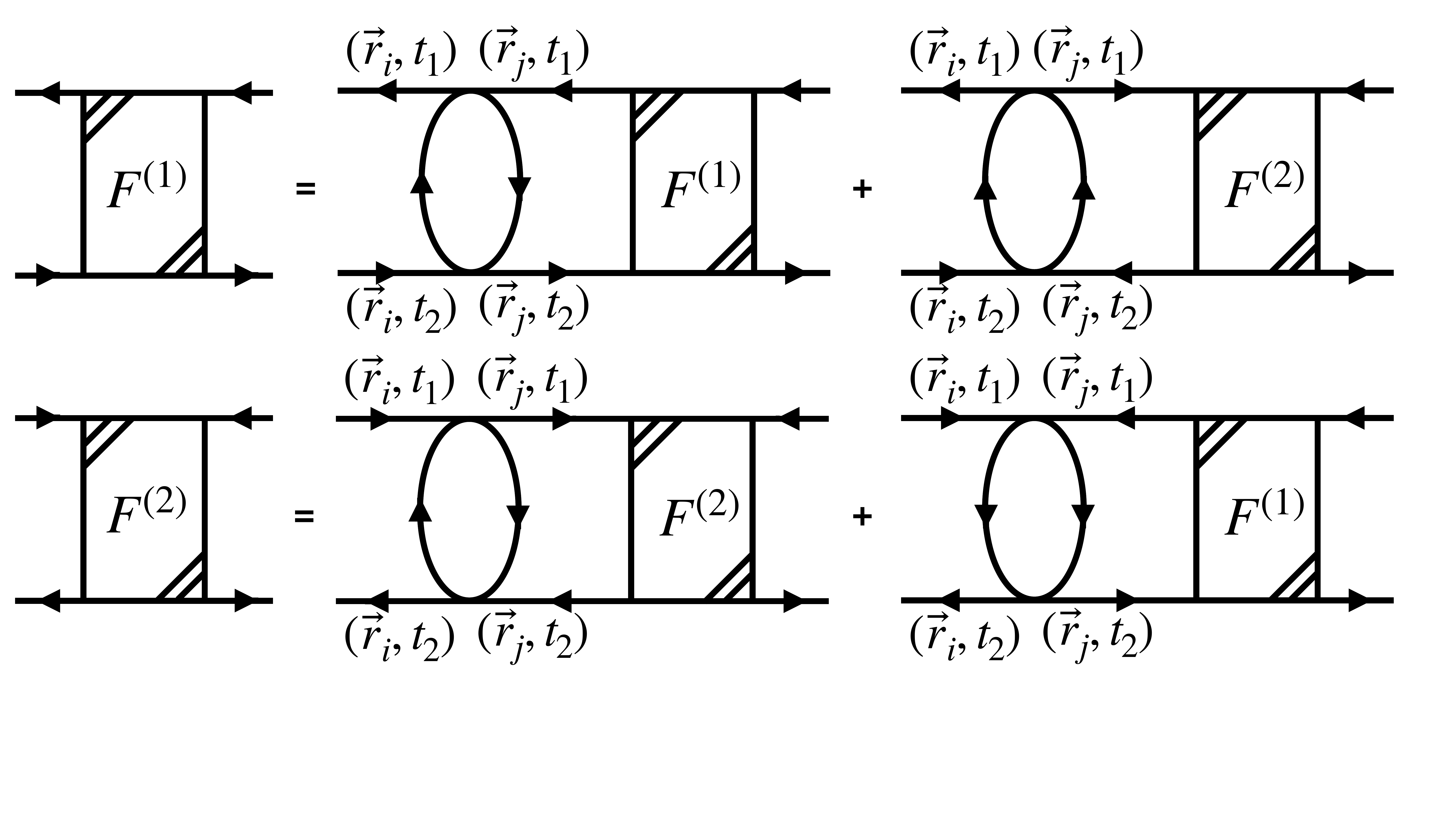}
\end{center}
\captionsetup{justification=RaggedRight, singlelinecheck=false}
\caption{Feynman diagrams for the two OTOCs, $F^{(1)}(t_1, \tau_2)$ and $F^{(2)}(t_1, t_2)$, with the upper and lower Keldysh contours separated by half the thermal cycle, $t = i \tfrac{\beta}{2}$. Similar to $\Gamma^{(1)}(\tau_1, \tau_2)$ and $\Gamma^{(2)}(\tau_1, \tau_2)$, the diagrams on the upper and lower left consisting of a particle-hole like bubble, $G^{(l)}_{w+}(t_1 - t_2) G^{(l)}_{w-}(t_2 - t_1)$, insertion on the rung, while the upper and lower right diagrams consist of a particle-particle like bubbles insertion, $G^{(l)}_{w+}(t_1 - t_2)^2$ and $G^{(l)}_{w-}(t_2 - t_1)^2$.}
\label{fig: OTOC ladders}
\end{figure}

The dynamics of quantum chaos and entanglement can be probed by the squared anti-commutators, $C_{I}(\vec{r}, t) = \langle | \{c^{\dg}(\vec{r}, t), c(0, 0) \}|^2 \rangle = 2 - 2 Re[F_{I}(\vec{r}, t, t)]$ and  $C_{II}(\vec{r}, t) = \langle | \{c(\vec{r}, t), c(0, 0) \}|^2 \rangle = 2 - 2 Re[F_{II}(\vec{r}, t, t)]$, and following Ref.~\cite{MaldacenaDouglasJHEP2016}, we analyze the following two disorder averaged regularized OTOCs, with $y^4 = \tfrac{1}{Z}e^{-\beta H}$.
\begin{subequations}
\ba
F_{I}(t_1, t_2) & = & \frac{1}{N} \sum_{\vec{r}} \overline{ \langle y c(\vec{r}, t_1) y c^{\dg}(0) y c^{\dg}(\vec{r, t_2}) y c(0) \rangle } \hspace{0.5cm} \\
F_{II}(t_1, t_2) & = & \frac{1}{N} \sum_{\vec{r}} \overline{ \langle y c^{\dg}(\vec{r}, t_1) y c^{\dg}(0) y c(\vec{r, t_2}) y c(0) \rangle } \hspace{0.5cm}
\ea
\label{eqn: OTOC I II}
\end{subequations}
Utilizing an augmented Keldysh formalism discussed in Ref.~\cite{AleinerAnnPhys2016}, we derive the OTOC, $F(t_1, t_2) = F^{(0)}(t_1, t_2) + \frac{1}{N} F^{(1)}(t_1, t_2) + O \bl \frac{1}{N^2} \br$. The $F^{(0)}(t_1, t_2)$ term comes from the disconnected contractions of local effective Green's functions. Similar to the scattering vertices for $\chi^{(l)}_{JJ}(\vec{q} = 0, i \Omega)$, there are particle-hole and particle-particle type bubble insertions into the rungs of the ladders diagrams for $F^{(1)}(t_1, t_2)$, as shown shown in Fig.~\ref{fig: OTOC ladders}. 

Note that we are now working in real time, $t$, as opposed to imaginary time, $\tau$, and the Green's functions describing the legs are $G^{(R/A)}(\vec{k}, i \omega)$. The upper and lower Keldysh contours are separated by $t = i \tfrac{\beta}{2}$; hence the bubble insertions involve the Wightman correlator $G^{(l)}_{w \pm}(t) = i  G^{(l)}(\tau = i t \pm \tfrac{\beta}{2})$, with the particle-hole and particle-particle type bubbles given by $G^{(l)}_{w+}(t) G^{(l)}_{w -}(-t)$ and $G^{(l)}_{w-}(t)^2$ respectively. Following a particle-hole symmetry argument similar to that applied to $\Gamma(i \Omega)$, it can be demonstrated that $F^{(1)}_{I}(t_1, t_2)$ and $F^{(1)}_{II}(t_1, t_2)$ have the same scattering kernels $K_{1}(t_1, t_2)$ and $K_{2}(t_1, t_2)$ . 
\begin{subequations}
\ba
F^{(1)}_{I}(t, t') & = & \int d t_1 d t_2  K_{1}(t_1, t_2) F^{(1)}_{I}(t_1, t_2) \cr
& & + K_{2}(t_1, t_2) F^{(1)}_{II}(t_1, t_2) \\
F^{(1)}_{II}(t, t') & = & \int d t_1 d t_2  K_{1}(t_1, t_2) F^{(1)}_{II}(t_1, t_2) \cr
& & + K_{2}(t_1, t_2) F^{(1)}_{I}(t_1, t_2) \\
K_{1}(t, t') & = & 2 J^2 \int \frac{d \vec{k}}{(2 \pi)^2} G^{(R)} (\vec{k}, t_1 - t) G^{(A)}(\vec{k}, t' - t_2) \cr
& & \times G^{(l)}_{w+}(t_1 - t_2) G^{(l)}_{w-}(t_2 - t_1) \\\
K_{2}(t, t') & = & -  J^2 \int \frac{d \vec{k}}{(2 \pi)^2} G^{(R)} (\vec{k}, t_2 - t') G^{(A)}(\vec{k}, t - t_1) \cr
& & \times G^{(l)}_{w-}(t_2 - t_1) G^{(l)}_{w-}(t_2 - t_1)
\ea
\label{eqn: OTOC integral eqns}
\end{subequations}
Hence, $F^{(1)}_{I}(t_1, t_2) = F^{(1)}_{II}(t_1, t_2) \equiv F^{(1)}(t_1, t_2)$, and Eq.~\ref{eqn: OTOC integral eqns} simplifies to a single integral equation. This can be transformed into a differential equation, and at zero temperature, it  gives the following solution (see SOM for details).
\ba
\label{eqn: OTOC 1}
F^{(1)}(\omega, \Omega) & = & \frac{\alpha_2}{ \left[1 - \frac{\alpha_1}{D^2} \frac{3}{2} \frac{\omega + i \frac{\Omega}{2}}{\omega - i \frac{\Omega}{2}} \ln \frac{M(\omega + \frac{\Omega}{2})}{M^{*}(\omega - \frac{\Omega}{2})} \right]}
\ea

We point out here that, $F^{(1)}(\omega, \Omega)$ is the Fourier transform of $F^{(1)}(t^{-}_{12} = \tfrac{t_1 - t_2}{2}, t^{+}_{12} = \tfrac{t_1 + t_2}{2})$, and we are only interested in the $t^{-}_{12} = 0$ and large $t^{+}_{12}$-limit. Therefore, to leading order, $F^{(1)}(t^{-}_{12} = 0, t^{+}_{12})$ grows linearly with time.
\ba
\label{eqn: OTOC 2}
F^{(1)}(t^{-}_{12} = 0, t^{+}_{12} = t) & = & \frac{8}{\sqrt{3}} \frac{D^2}{J} t e^{- i \frac{16}{\sqrt{3}} \frac{D^2}{J} t}
\ea

As discussed in Ref.~\cite{ZhaiSciBul2017}, the second R\'{e}nyi entropy, $S^{(2)}_A$, of the system is related to the OTOC; hence,  the $t$-linear envelope in Eq.~\ref{eqn: OTOC 2} confirms that the entanglement entropy of the system grows logarithmically with time, thereby verifying that the system is in an MBL ground state. This is consistent with numerical studies of the 1-D random field XXZ model \cite{HusePalPRB2010, MoorePRL2012, AbaninSerbynPRX2015, ZhaiSciBul2017}, as well as recent numerical studies of generalized SYK models that have found an MBL phase by analyzing the level statistics \cite{YaoJianPRL2017, TezukaGarciaPRL2018}. As discussed in Ref.~\cite{FradkinHuseAnnPhys2017}, the OTOC has an oscillatory component due to the exponentially decaying effective interactions, and de-phasing between different eigenstates results in the power-law envelope. The oscillatory component in Eq.~\ref{eqn: OTOC 2} results from the disorder-averaged high-energy incoherent background described by the logarithmic term in $\Sigma(i \omega_n)$; hence, it oscillates at a frequency $\tfrac{D^2}{J}$ set by the incoherent background.

Disorder averaging of the SYK interactions results in a strictly local interaction at the large-$N$ saddle point; hence, we are only able to analyze the uniform component of the OTOC. It is therefore instructive to examine the bubble diagram for the OTOC, $F^{(b)}(\vec{r}, t_1, t_2)$, which will provide insight into the real-space $\vec{r}$-dependence of the OTOC (refer to SOM for details of calculations). The squared commutator is proportional to the real part of the OTOC, and we obtain,
\ba
\label{eqn: OTOC bubble}
\Re[F^{(b)} (\vec{r}, t^{-}_{12} = 0,t^{+}_{12} = t)] & = & \frac{2 D^2}{\pi^{5/2} J}  \frac{e^{- \frac{\pi |\vec{r}|}{a}}}{ (v_F r)^2} \ln |t| \hspace{0.3cm}
\ea

We note that the logarithmic dependence on $t$ is the origin of the $t$-linear behavior of $F^{1}(t^{-}_{12} = 0, t^{+}_{12} = t)$, resulting from summing up the infinite series of the ladder diagrams. In addition, we note that it is exponentially localized with a localization length of $a$, the lattice spacing, which is expected given that the SYK interactions drive the formation of localized states. Hence, this provides an insight into the exponentially weak interactions that drive entanglement in the MBL phase\cite{HusePalPRB2010, MoorePRL2012, AltmanVoskPRL2013, AbaninPRL2013}. $\Re[F^{(b)} (\vec{r}, t^{-}_{12} = 0,t^{+}_{12} = t)]$ also has an oscillatory component, similarly due to disorder-averaging of the high-energy background, which is relegated to the SOM for convenience.

In conclusion, by making use of an asymptotic solution of the large-N saddle point equations, we have found that a two-dimensional itinerant SYK model of spin-less fermions forms a many-body localized phase in the strong coupling regime. The system forms a lattice of interaction-driven bound states, with a high-energy background, and the ground state is verified to be insulating by calculating the current-current correlation function, which is found to be zero in the small $\omega$-limit. The many-body-localized nature of the ground state is confirmed by calculating the OTOC, and we find that the OTOC has a linear in time behavior, meaning that the entanglement entropy of the system grows logarithmically with time; thereby confirming the system forms a many-body localized phase.

We gratefully acknowledge N. Nagaosa, P. Coleman and N. Tsuji for many enlightening discussions. This work was supported by the Japan Science and Technology Agency (JST), CREST Grant No. JPMJCR16F1.

\bibliography{SYK}

\begin{thebibliography}{45}%
\makeatletter
\providecommand \@ifxundefined [1]{%
 \@ifx{#1\undefined}
}%
\providecommand \@ifnum [1]{%
 \ifnum #1\expandafter \@firstoftwo
 \else \expandafter \@secondoftwo
 \fi
}%
\providecommand \@ifx [1]{%
 \ifx #1\expandafter \@firstoftwo
 \else \expandafter \@secondoftwo
 \fi
}%
\providecommand \natexlab [1]{#1}%
\providecommand \enquote  [1]{``#1''}%
\providecommand \bibnamefont  [1]{#1}%
\providecommand \bibfnamefont [1]{#1}%
\providecommand \citenamefont [1]{#1}%
\providecommand \href@noop [0]{\@secondoftwo}%
\providecommand \href [0]{\begingroup \@sanitize@url \@href}%
\providecommand \@href[1]{\@@startlink{#1}\@@href}%
\providecommand \@@href[1]{\endgroup#1\@@endlink}%
\providecommand \@sanitize@url [0]{\catcode `\\12\catcode `\$12\catcode
  `\&12\catcode `\#12\catcode `\^12\catcode `\_12\catcode `\%12\relax}%
\providecommand \@@startlink[1]{}%
\providecommand \@@endlink[0]{}%
\providecommand \url  [0]{\begingroup\@sanitize@url \@url }%
\providecommand \@url [1]{\endgroup\@href {#1}{\urlprefix }}%
\providecommand \urlprefix  [0]{URL }%
\providecommand \Eprint [0]{\href }%
\providecommand \doibase [0]{http://dx.doi.org/}%
\providecommand \selectlanguage [0]{\@gobble}%
\providecommand \bibinfo  [0]{\@secondoftwo}%
\providecommand \bibfield  [0]{\@secondoftwo}%
\providecommand \translation [1]{[#1]}%
\providecommand \BibitemOpen [0]{}%
\providecommand \bibitemStop [0]{}%
\providecommand \bibitemNoStop [0]{.\EOS\space}%
\providecommand \EOS [0]{\spacefactor3000\relax}%
\providecommand \BibitemShut  [1]{\csname bibitem#1\endcsname}%
\let\auto@bib@innerbib\@empty
%</preamble>
\bibitem [{\citenamefont {Anderson}(1958)}]{AndersonPhysRev1958}%
  \BibitemOpen
  \bibfield  {author} {\bibinfo {author} {\bibfnamefont {P.~W.}\ \bibnamefont
  {Anderson}},\ }\href {\doibase 10.1103/PhysRev.109.1492} {\bibfield
  {journal} {\bibinfo  {journal} {Phys. Rev.}\ }\textbf {\bibinfo {volume}
  {109}},\ \bibinfo {pages} {1492} (\bibinfo {year} {1958})}\BibitemShut
  {NoStop}%
\bibitem [{\citenamefont {Abrahams}\ \emph {et~al.}(1979)\citenamefont
  {Abrahams}, \citenamefont {Anderson}, \citenamefont {Licciardello},\ and\
  \citenamefont {Ramakrishnan}}]{AbrahamsAndersonPRL1979}%
  \BibitemOpen
  \bibfield  {author} {\bibinfo {author} {\bibfnamefont {E.}~\bibnamefont
  {Abrahams}}, \bibinfo {author} {\bibfnamefont {P.~W.}\ \bibnamefont
  {Anderson}}, \bibinfo {author} {\bibfnamefont {D.~C.}\ \bibnamefont
  {Licciardello}}, \ and\ \bibinfo {author} {\bibfnamefont {T.~V.}\
  \bibnamefont {Ramakrishnan}},\ }\href {\doibase 10.1103/PhysRevLett.42.673}
  {\bibfield  {journal} {\bibinfo  {journal} {Phys. Rev. Lett.}\ }\textbf
  {\bibinfo {volume} {42}},\ \bibinfo {pages} {673} (\bibinfo {year}
  {1979})}\BibitemShut {NoStop}%
\bibitem [{\citenamefont {Evers}\ and\ \citenamefont
  {Mirlin}(2008)}]{MirlinRMP2008}%
  \BibitemOpen
  \bibfield  {author} {\bibinfo {author} {\bibfnamefont {F.}~\bibnamefont
  {Evers}}\ and\ \bibinfo {author} {\bibfnamefont {A.~D.}\ \bibnamefont
  {Mirlin}},\ }\href {\doibase 10.1103/RevModPhys.80.1355} {\bibfield
  {journal} {\bibinfo  {journal} {Rev. Mod. Phys.}\ }\textbf {\bibinfo {volume}
  {80}},\ \bibinfo {pages} {1355} (\bibinfo {year} {2008})}\BibitemShut
  {NoStop}%
\bibitem [{\citenamefont {Basko}\ \emph {et~al.}(2006)\citenamefont {Basko},
  \citenamefont {Aleiner},\ and\ \citenamefont
  {Altshuler}}]{AltshulerAnnPhys2006}%
  \BibitemOpen
  \bibfield  {author} {\bibinfo {author} {\bibfnamefont {D.}~\bibnamefont
  {Basko}}, \bibinfo {author} {\bibfnamefont {I.}~\bibnamefont {Aleiner}}, \
  and\ \bibinfo {author} {\bibfnamefont {B.}~\bibnamefont {Altshuler}},\ }\href
  {\doibase https://doi.org/10.1016/j.aop.2005.11.014} {\bibfield  {journal}
  {\bibinfo  {journal} {Annals of Physics}\ }\textbf {\bibinfo {volume}
  {321}},\ \bibinfo {pages} {1126 } (\bibinfo {year} {2006})}\BibitemShut
  {NoStop}%
\bibitem [{\citenamefont {Serbyn}\ \emph {et~al.}(2013)\citenamefont {Serbyn},
  \citenamefont {Papi\ifmmode~\acute{c}\else \'{c}\fi{}},\ and\ \citenamefont
  {Abanin}}]{AbaninPRL2013}%
  \BibitemOpen
  \bibfield  {author} {\bibinfo {author} {\bibfnamefont {M.}~\bibnamefont
  {Serbyn}}, \bibinfo {author} {\bibfnamefont {Z.}~\bibnamefont
  {Papi\ifmmode~\acute{c}\else \'{c}\fi{}}}, \ and\ \bibinfo {author}
  {\bibfnamefont {D.~A.}\ \bibnamefont {Abanin}},\ }\href {\doibase
  10.1103/PhysRevLett.110.260601} {\bibfield  {journal} {\bibinfo  {journal}
  {Phys. Rev. Lett.}\ }\textbf {\bibinfo {volume} {110}},\ \bibinfo {pages}
  {260601} (\bibinfo {year} {2013})}\BibitemShut {NoStop}%
\bibitem [{\citenamefont {Huse}\ \emph {et~al.}(2014)\citenamefont {Huse},
  \citenamefont {Nandkishore},\ and\ \citenamefont {Oganesyan}}]{HusePRB2014}%
  \BibitemOpen
  \bibfield  {author} {\bibinfo {author} {\bibfnamefont {D.~A.}\ \bibnamefont
  {Huse}}, \bibinfo {author} {\bibfnamefont {R.}~\bibnamefont {Nandkishore}}, \
  and\ \bibinfo {author} {\bibfnamefont {V.}~\bibnamefont {Oganesyan}},\ }\href
  {\doibase 10.1103/PhysRevB.90.174202} {\bibfield  {journal} {\bibinfo
  {journal} {Phys. Rev. B}\ }\textbf {\bibinfo {volume} {90}},\ \bibinfo
  {pages} {174202} (\bibinfo {year} {2014})}\BibitemShut {NoStop}%
\bibitem [{\citenamefont {Huse}\ \emph {et~al.}(2013)\citenamefont {Huse},
  \citenamefont {Nandkishore}, \citenamefont {Oganesyan}, \citenamefont {Pal},\
  and\ \citenamefont {Sondhi}}]{SondhiHusePRB2013}%
  \BibitemOpen
  \bibfield  {author} {\bibinfo {author} {\bibfnamefont {D.~A.}\ \bibnamefont
  {Huse}}, \bibinfo {author} {\bibfnamefont {R.}~\bibnamefont {Nandkishore}},
  \bibinfo {author} {\bibfnamefont {V.}~\bibnamefont {Oganesyan}}, \bibinfo
  {author} {\bibfnamefont {A.}~\bibnamefont {Pal}}, \ and\ \bibinfo {author}
  {\bibfnamefont {S.~L.}\ \bibnamefont {Sondhi}},\ }\href {\doibase
  10.1103/PhysRevB.88.014206} {\bibfield  {journal} {\bibinfo  {journal} {Phys.
  Rev. B}\ }\textbf {\bibinfo {volume} {88}},\ \bibinfo {pages} {014206}
  (\bibinfo {year} {2013})}\BibitemShut {NoStop}%
\bibitem [{\citenamefont {Imbrie}(2016)}]{ImbrieJStatPhys2016}%
  \BibitemOpen
  \bibfield  {author} {\bibinfo {author} {\bibfnamefont {J.~Z.}\ \bibnamefont
  {Imbrie}},\ }\href {\doibase 10.1007/s10955-016-1508-x} {\bibfield  {journal}
  {\bibinfo  {journal} {Journal of Statistical Physics}\ }\textbf {\bibinfo
  {volume} {163}},\ \bibinfo {pages} {998} (\bibinfo {year}
  {2016})}\BibitemShut {NoStop}%
\bibitem [{\citenamefont {Deutsch}(1991)}]{DeutschPRA1991}%
  \BibitemOpen
  \bibfield  {author} {\bibinfo {author} {\bibfnamefont {J.~M.}\ \bibnamefont
  {Deutsch}},\ }\href {\doibase 10.1103/PhysRevA.43.2046} {\bibfield  {journal}
  {\bibinfo  {journal} {Phys. Rev. A}\ }\textbf {\bibinfo {volume} {43}},\
  \bibinfo {pages} {2046} (\bibinfo {year} {1991})}\BibitemShut {NoStop}%
\bibitem [{\citenamefont {Srednicki}(1994)}]{SrednickiPRE1994}%
  \BibitemOpen
  \bibfield  {author} {\bibinfo {author} {\bibfnamefont {M.}~\bibnamefont
  {Srednicki}},\ }\href {\doibase 10.1103/PhysRevE.50.888} {\bibfield
  {journal} {\bibinfo  {journal} {Phys. Rev. E}\ }\textbf {\bibinfo {volume}
  {50}},\ \bibinfo {pages} {888} (\bibinfo {year} {1994})}\BibitemShut
  {NoStop}%
\bibitem [{\citenamefont {Rigol}\ \emph {et~al.}(2008)\citenamefont {Rigol},
  \citenamefont {Dunjko},\ and\ \citenamefont {Olshanii}}]{OlshaniiNature2008}%
  \BibitemOpen
  \bibfield  {author} {\bibinfo {author} {\bibfnamefont {M.}~\bibnamefont
  {Rigol}}, \bibinfo {author} {\bibfnamefont {V.}~\bibnamefont {Dunjko}}, \
  and\ \bibinfo {author} {\bibfnamefont {M.}~\bibnamefont {Olshanii}},\ }\href
  {https://doi.org/10.1038/nature06838} {\bibfield  {journal} {\bibinfo
  {journal} {Nature}\ }\textbf {\bibinfo {volume} {452}},\ \bibinfo {pages}
  {854 EP } (\bibinfo {year} {2008})}\BibitemShut {NoStop}%
\bibitem [{\citenamefont {Chiara}\ \emph {et~al.}(2006)\citenamefont {Chiara},
  \citenamefont {Montangero}, \citenamefont {Calabrese},\ and\ \citenamefont
  {Fazio}}]{FazioJStatMech2006}%
  \BibitemOpen
  \bibfield  {author} {\bibinfo {author} {\bibfnamefont {G.~D.}\ \bibnamefont
  {Chiara}}, \bibinfo {author} {\bibfnamefont {S.}~\bibnamefont {Montangero}},
  \bibinfo {author} {\bibfnamefont {P.}~\bibnamefont {Calabrese}}, \ and\
  \bibinfo {author} {\bibfnamefont {R.}~\bibnamefont {Fazio}},\ }\href
  {\doibase 10.1088/1742-5468/2006/03/p03001} {\bibfield  {journal} {\bibinfo
  {journal} {Journal of Statistical Mechanics: Theory and Experiment}\ }\textbf
  {\bibinfo {volume} {2006}},\ \bibinfo {pages} {P03001} (\bibinfo {year}
  {2006})}\BibitemShut {NoStop}%
\bibitem [{\citenamefont {\ifmmode \check{Z}\else
  \v{Z}\fi{}nidari\ifmmode~\check{c}\else \v{c}\fi{}}\ \emph
  {et~al.}(2008)\citenamefont {\ifmmode \check{Z}\else
  \v{Z}\fi{}nidari\ifmmode~\check{c}\else \v{c}\fi{}}, \citenamefont {Prosen},\
  and\ \citenamefont {Prelov\ifmmode~\check{s}\else
  \v{s}\fi{}ek}}]{PrelovsekPRB2008}%
  \BibitemOpen
  \bibfield  {author} {\bibinfo {author} {\bibfnamefont {M.}~\bibnamefont
  {\ifmmode \check{Z}\else \v{Z}\fi{}nidari\ifmmode~\check{c}\else
  \v{c}\fi{}}}, \bibinfo {author} {\bibfnamefont {T.~c.~v.}\ \bibnamefont
  {Prosen}}, \ and\ \bibinfo {author} {\bibfnamefont {P.}~\bibnamefont
  {Prelov\ifmmode~\check{s}\else \v{s}\fi{}ek}},\ }\href {\doibase
  10.1103/PhysRevB.77.064426} {\bibfield  {journal} {\bibinfo  {journal} {Phys.
  Rev. B}\ }\textbf {\bibinfo {volume} {77}},\ \bibinfo {pages} {064426}
  (\bibinfo {year} {2008})}\BibitemShut {NoStop}%
\bibitem [{\citenamefont {Bardarson}\ \emph {et~al.}(2012)\citenamefont
  {Bardarson}, \citenamefont {Pollmann},\ and\ \citenamefont
  {Moore}}]{MoorePRL2012}%
  \BibitemOpen
  \bibfield  {author} {\bibinfo {author} {\bibfnamefont {J.~H.}\ \bibnamefont
  {Bardarson}}, \bibinfo {author} {\bibfnamefont {F.}~\bibnamefont {Pollmann}},
  \ and\ \bibinfo {author} {\bibfnamefont {J.~E.}\ \bibnamefont {Moore}},\
  }\href {\doibase 10.1103/PhysRevLett.109.017202} {\bibfield  {journal}
  {\bibinfo  {journal} {Phys. Rev. Lett.}\ }\textbf {\bibinfo {volume} {109}},\
  \bibinfo {pages} {017202} (\bibinfo {year} {2012})}\BibitemShut {NoStop}%
\bibitem [{\citenamefont {Huang}\ \emph {et~al.}(2017)\citenamefont {Huang},
  \citenamefont {Zhang},\ and\ \citenamefont {Chen}}]{ChenAnnPhys2016}%
  \BibitemOpen
  \bibfield  {author} {\bibinfo {author} {\bibfnamefont {Y.}~\bibnamefont
  {Huang}}, \bibinfo {author} {\bibfnamefont {Y.-L.}\ \bibnamefont {Zhang}}, \
  and\ \bibinfo {author} {\bibfnamefont {X.}~\bibnamefont {Chen}},\ }\href
  {\doibase 10.1002/andp.201600318} {\bibfield  {journal} {\bibinfo  {journal}
  {Annalen der Physik}\ }\textbf {\bibinfo {volume} {529}},\ \bibinfo {pages}
  {1600318} (\bibinfo {year} {2017})}\BibitemShut {NoStop}%
\bibitem [{\citenamefont {Fan}\ \emph {et~al.}(2017)\citenamefont {Fan},
  \citenamefont {Zhang}, \citenamefont {Shen},\ and\ \citenamefont
  {Zhai}}]{ZhaiSciBul2017}%
  \BibitemOpen
  \bibfield  {author} {\bibinfo {author} {\bibfnamefont {R.}~\bibnamefont
  {Fan}}, \bibinfo {author} {\bibfnamefont {P.}~\bibnamefont {Zhang}}, \bibinfo
  {author} {\bibfnamefont {H.}~\bibnamefont {Shen}}, \ and\ \bibinfo {author}
  {\bibfnamefont {H.}~\bibnamefont {Zhai}},\ }\href {\doibase
  https://doi.org/10.1016/j.scib.2017.04.011} {\bibfield  {journal} {\bibinfo
  {journal} {Science Bulletin}\ }\textbf {\bibinfo {volume} {62}},\ \bibinfo
  {pages} {707 } (\bibinfo {year} {2017})}\BibitemShut {NoStop}%
\bibitem [{\citenamefont {Abdul-Rahman}\ \emph {et~al.}(2016)\citenamefont
  {Abdul-Rahman}, \citenamefont {Nachtergaele}, \citenamefont {Sims},\ and\
  \citenamefont {Stolz}}]{StolzLettMathPhys2016}%
  \BibitemOpen
  \bibfield  {author} {\bibinfo {author} {\bibfnamefont {H.}~\bibnamefont
  {Abdul-Rahman}}, \bibinfo {author} {\bibfnamefont {B.}~\bibnamefont
  {Nachtergaele}}, \bibinfo {author} {\bibfnamefont {R.}~\bibnamefont {Sims}},
  \ and\ \bibinfo {author} {\bibfnamefont {G.}~\bibnamefont {Stolz}},\ }\href
  {\doibase 10.1007/s11005-016-0835-9} {\bibfield  {journal} {\bibinfo
  {journal} {Letters in Mathematical Physics}\ }\textbf {\bibinfo {volume}
  {106}},\ \bibinfo {pages} {649} (\bibinfo {year} {2016})}\BibitemShut
  {NoStop}%
\bibitem [{\citenamefont {Kim}\ and\ \citenamefont {Huse}(2013)}]{HusePRL2013}%
  \BibitemOpen
  \bibfield  {author} {\bibinfo {author} {\bibfnamefont {H.}~\bibnamefont
  {Kim}}\ and\ \bibinfo {author} {\bibfnamefont {D.~A.}\ \bibnamefont {Huse}},\
  }\href {\doibase 10.1103/PhysRevLett.111.127205} {\bibfield  {journal}
  {\bibinfo  {journal} {Phys. Rev. Lett.}\ }\textbf {\bibinfo {volume} {111}},\
  \bibinfo {pages} {127205} (\bibinfo {year} {2013})}\BibitemShut {NoStop}%
\bibitem [{\citenamefont {Schachenmayer}\ \emph {et~al.}(2013)\citenamefont
  {Schachenmayer}, \citenamefont {Lanyon}, \citenamefont {Roos},\ and\
  \citenamefont {Daley}}]{DaleyPRX2013}%
  \BibitemOpen
  \bibfield  {author} {\bibinfo {author} {\bibfnamefont {J.}~\bibnamefont
  {Schachenmayer}}, \bibinfo {author} {\bibfnamefont {B.~P.}\ \bibnamefont
  {Lanyon}}, \bibinfo {author} {\bibfnamefont {C.~F.}\ \bibnamefont {Roos}}, \
  and\ \bibinfo {author} {\bibfnamefont {A.~J.}\ \bibnamefont {Daley}},\ }\href
  {\doibase 10.1103/PhysRevX.3.031015} {\bibfield  {journal} {\bibinfo
  {journal} {Phys. Rev. X}\ }\textbf {\bibinfo {volume} {3}},\ \bibinfo {pages}
  {031015} (\bibinfo {year} {2013})}\BibitemShut {NoStop}%
\bibitem [{\citenamefont {Larkin}\ and\ \citenamefont
  {Ovchinnikov}(1969)}]{LarkinOvchinnikovJETP1969}%
  \BibitemOpen
  \bibfield  {author} {\bibinfo {author} {\bibfnamefont {A.~I.}\ \bibnamefont
  {Larkin}}\ and\ \bibinfo {author} {\bibfnamefont {Y.~N.}\ \bibnamefont
  {Ovchinnikov}},\ }\href@noop {} {\bibfield  {journal} {\bibinfo  {journal}
  {Sov. Phys. JETP}\ }\textbf {\bibinfo {volume} {28}},\ \bibinfo {pages}
  {1200} (\bibinfo {year} {1969})}\BibitemShut {NoStop}%
\bibitem [{\citenamefont {Kitaev}(2014)}]{KitaevTalk2014}%
  \BibitemOpen
  \bibfield  {author} {\bibinfo {author} {\bibfnamefont {A.}~\bibnamefont
  {Kitaev}},\ }\href@noop {} {\enquote {\bibinfo {title} {Fundamental physics
  prize symposium talk},}\ } (\bibinfo {year} {2014})\BibitemShut {NoStop}%
\bibitem [{\citenamefont {Kitaev}(2015)}]{KitaevKITP2015}%
  \BibitemOpen
  \bibfield  {author} {\bibinfo {author} {\bibfnamefont {A.}~\bibnamefont
  {Kitaev}},\ }\href@noop {} {\enquote {\bibinfo {title} {Kitp program:
  Entanglement in strongly-correlated quantum matter, 2015},}\ } (\bibinfo
  {year} {2015}),\ \bibinfo {note}
  {http://online.kitp.ucsb.edu/online/entangled15/kitaev}\BibitemShut {NoStop}%
\bibitem [{\citenamefont {Shenker}\ and\ \citenamefont
  {Stanford}(2015)}]{ShenkerStanfordJHEP2015}%
  \BibitemOpen
  \bibfield  {author} {\bibinfo {author} {\bibfnamefont {S.~H.}\ \bibnamefont
  {Shenker}}\ and\ \bibinfo {author} {\bibfnamefont {D.}~\bibnamefont
  {Stanford}},\ }\href {\doibase 10.1007/JHEP05(2015)132} {\bibfield  {journal}
  {\bibinfo  {journal} {Journal of High Energy Physics}\ }\textbf {\bibinfo
  {volume} {2015}},\ \bibinfo {pages} {132} (\bibinfo {year}
  {2015})}\BibitemShut {NoStop}%
\bibitem [{\citenamefont {Maldacena}\ \emph
  {et~al.}(2016{\natexlab{a}})\citenamefont {Maldacena}, \citenamefont
  {Shenker},\ and\ \citenamefont {Stanford}}]{MaldacenaStanfordJHEP2016}%
  \BibitemOpen
  \bibfield  {author} {\bibinfo {author} {\bibfnamefont {J.}~\bibnamefont
  {Maldacena}}, \bibinfo {author} {\bibfnamefont {S.~H.}\ \bibnamefont
  {Shenker}}, \ and\ \bibinfo {author} {\bibfnamefont {D.}~\bibnamefont
  {Stanford}},\ }\href {\doibase 10.1007/JHEP08(2016)106} {\bibfield  {journal}
  {\bibinfo  {journal} {Journal of High Energy Physics}\ }\textbf {\bibinfo
  {volume} {2016}},\ \bibinfo {pages} {106} (\bibinfo {year}
  {2016}{\natexlab{a}})}\BibitemShut {NoStop}%
\bibitem [{\citenamefont {Maldacena}\ and\ \citenamefont
  {Stanford}(2016)}]{MaldacenaStanfordPRD2016}%
  \BibitemOpen
  \bibfield  {author} {\bibinfo {author} {\bibfnamefont {J.}~\bibnamefont
  {Maldacena}}\ and\ \bibinfo {author} {\bibfnamefont {D.}~\bibnamefont
  {Stanford}},\ }\href {\doibase 10.1103/PhysRevD.94.106002} {\bibfield
  {journal} {\bibinfo  {journal} {Phys. Rev. D}\ }\textbf {\bibinfo {volume}
  {94}},\ \bibinfo {pages} {106002} (\bibinfo {year} {2016})}\BibitemShut
  {NoStop}%
\bibitem [{\citenamefont {Hosur}\ \emph {et~al.}(2016)\citenamefont {Hosur},
  \citenamefont {Qi}, \citenamefont {Roberts},\ and\ \citenamefont
  {Yoshida}}]{YoshidaQiJHEP2016}%
  \BibitemOpen
  \bibfield  {author} {\bibinfo {author} {\bibfnamefont {P.}~\bibnamefont
  {Hosur}}, \bibinfo {author} {\bibfnamefont {X.-L.}\ \bibnamefont {Qi}},
  \bibinfo {author} {\bibfnamefont {D.~A.}\ \bibnamefont {Roberts}}, \ and\
  \bibinfo {author} {\bibfnamefont {B.}~\bibnamefont {Yoshida}},\ }\href
  {\doibase 10.1007/JHEP02(2016)004} {\bibfield  {journal} {\bibinfo  {journal}
  {Journal of High Energy Physics}\ }\textbf {\bibinfo {volume} {2016}},\
  \bibinfo {pages} {4} (\bibinfo {year} {2016})}\BibitemShut {NoStop}%
\bibitem [{\citenamefont {Sachdev}\ and\ \citenamefont
  {Ye}(1993)}]{SachdevYePRL1993}%
  \BibitemOpen
  \bibfield  {author} {\bibinfo {author} {\bibfnamefont {S.}~\bibnamefont
  {Sachdev}}\ and\ \bibinfo {author} {\bibfnamefont {J.}~\bibnamefont {Ye}},\
  }\href {\doibase 10.1103/PhysRevLett.70.3339} {\bibfield  {journal} {\bibinfo
   {journal} {Phys. Rev. Lett.}\ }\textbf {\bibinfo {volume} {70}},\ \bibinfo
  {pages} {3339} (\bibinfo {year} {1993})}\BibitemShut {NoStop}%
\bibitem [{\citenamefont {Sachdev}(2015)}]{SachdevPRX2015}%
  \BibitemOpen
  \bibfield  {author} {\bibinfo {author} {\bibfnamefont {S.}~\bibnamefont
  {Sachdev}},\ }\href {\doibase 10.1103/PhysRevX.5.041025} {\bibfield
  {journal} {\bibinfo  {journal} {Phys. Rev. X}\ }\textbf {\bibinfo {volume}
  {5}},\ \bibinfo {pages} {041025} (\bibinfo {year} {2015})}\BibitemShut
  {NoStop}%
\bibitem [{\citenamefont {Davison}\ \emph {et~al.}(2017)\citenamefont
  {Davison}, \citenamefont {Fu}, \citenamefont {Georges}, \citenamefont {Gu},
  \citenamefont {Jensen},\ and\ \citenamefont
  {Sachdev}}]{SachdevGeorgesPRB2017}%
  \BibitemOpen
  \bibfield  {author} {\bibinfo {author} {\bibfnamefont {R.~A.}\ \bibnamefont
  {Davison}}, \bibinfo {author} {\bibfnamefont {W.}~\bibnamefont {Fu}},
  \bibinfo {author} {\bibfnamefont {A.}~\bibnamefont {Georges}}, \bibinfo
  {author} {\bibfnamefont {Y.}~\bibnamefont {Gu}}, \bibinfo {author}
  {\bibfnamefont {K.}~\bibnamefont {Jensen}}, \ and\ \bibinfo {author}
  {\bibfnamefont {S.}~\bibnamefont {Sachdev}},\ }\href {\doibase
  10.1103/PhysRevB.95.155131} {\bibfield  {journal} {\bibinfo  {journal} {Phys.
  Rev. B}\ }\textbf {\bibinfo {volume} {95}},\ \bibinfo {pages} {155131}
  (\bibinfo {year} {2017})}\BibitemShut {NoStop}%
\bibitem [{\citenamefont {Gu}\ \emph {et~al.}(2017{\natexlab{a}})\citenamefont
  {Gu}, \citenamefont {Lucas},\ and\ \citenamefont {Qi}}]{QiGuSciPost2017}%
  \BibitemOpen
  \bibfield  {author} {\bibinfo {author} {\bibfnamefont {Y.}~\bibnamefont
  {Gu}}, \bibinfo {author} {\bibfnamefont {A.}~\bibnamefont {Lucas}}, \ and\
  \bibinfo {author} {\bibfnamefont {X.-L.}\ \bibnamefont {Qi}},\ }\href
  {\doibase 10.21468/SciPostPhys.2.3.018} {\bibfield  {journal} {\bibinfo
  {journal} {SciPost Phys.}\ }\textbf {\bibinfo {volume} {2}},\ \bibinfo
  {pages} {018} (\bibinfo {year} {2017}{\natexlab{a}})}\BibitemShut {NoStop}%
\bibitem [{\citenamefont {Gu}\ \emph {et~al.}(2017{\natexlab{b}})\citenamefont
  {Gu}, \citenamefont {Qi},\ and\ \citenamefont
  {Stanford}}]{StanfordQiJHEP2017}%
  \BibitemOpen
  \bibfield  {author} {\bibinfo {author} {\bibfnamefont {Y.}~\bibnamefont
  {Gu}}, \bibinfo {author} {\bibfnamefont {X.-L.}\ \bibnamefont {Qi}}, \ and\
  \bibinfo {author} {\bibfnamefont {D.}~\bibnamefont {Stanford}},\ }\href
  {\doibase 10.1007/JHEP05(2017)125} {\bibfield  {journal} {\bibinfo  {journal}
  {Journal of High Energy Physics}\ }\textbf {\bibinfo {volume} {2017}},\
  \bibinfo {pages} {125} (\bibinfo {year} {2017}{\natexlab{b}})}\BibitemShut
  {NoStop}%
\bibitem [{\citenamefont {You}\ \emph {et~al.}(2017)\citenamefont {You},
  \citenamefont {Ludwig},\ and\ \citenamefont {Xu}}]{XuLudwigPRB2017}%
  \BibitemOpen
  \bibfield  {author} {\bibinfo {author} {\bibfnamefont {Y.-Z.}\ \bibnamefont
  {You}}, \bibinfo {author} {\bibfnamefont {A.~W.~W.}\ \bibnamefont {Ludwig}},
  \ and\ \bibinfo {author} {\bibfnamefont {C.}~\bibnamefont {Xu}},\ }\href
  {\doibase 10.1103/PhysRevB.95.115150} {\bibfield  {journal} {\bibinfo
  {journal} {Phys. Rev. B}\ }\textbf {\bibinfo {volume} {95}},\ \bibinfo
  {pages} {115150} (\bibinfo {year} {2017})}\BibitemShut {NoStop}%
\bibitem [{\citenamefont {Chen}\ \emph
  {et~al.}(2017{\natexlab{a}})\citenamefont {Chen}, \citenamefont {Fan},
  \citenamefont {Chen}, \citenamefont {Zhai},\ and\ \citenamefont
  {Zhang}}]{ZhangChenPRL2017}%
  \BibitemOpen
  \bibfield  {author} {\bibinfo {author} {\bibfnamefont {X.}~\bibnamefont
  {Chen}}, \bibinfo {author} {\bibfnamefont {R.}~\bibnamefont {Fan}}, \bibinfo
  {author} {\bibfnamefont {Y.}~\bibnamefont {Chen}}, \bibinfo {author}
  {\bibfnamefont {H.}~\bibnamefont {Zhai}}, \ and\ \bibinfo {author}
  {\bibfnamefont {P.}~\bibnamefont {Zhang}},\ }\href {\doibase
  10.1103/PhysRevLett.119.207603} {\bibfield  {journal} {\bibinfo  {journal}
  {Phys. Rev. Lett.}\ }\textbf {\bibinfo {volume} {119}},\ \bibinfo {pages}
  {207603} (\bibinfo {year} {2017}{\natexlab{a}})}\BibitemShut {NoStop}%
\bibitem [{\citenamefont {Banerjee}\ and\ \citenamefont
  {Altman}(2017)}]{AltmanPRB2017}%
  \BibitemOpen
  \bibfield  {author} {\bibinfo {author} {\bibfnamefont {S.}~\bibnamefont
  {Banerjee}}\ and\ \bibinfo {author} {\bibfnamefont {E.}~\bibnamefont
  {Altman}},\ }\href {\doibase 10.1103/PhysRevB.95.134302} {\bibfield
  {journal} {\bibinfo  {journal} {Phys. Rev. B}\ }\textbf {\bibinfo {volume}
  {95}},\ \bibinfo {pages} {134302} (\bibinfo {year} {2017})}\BibitemShut
  {NoStop}%
\bibitem [{\citenamefont {Liu}\ \emph {et~al.}(2018)\citenamefont {Liu},
  \citenamefont {Chen},\ and\ \citenamefont {Balents}}]{BalentsPRB2018}%
  \BibitemOpen
  \bibfield  {author} {\bibinfo {author} {\bibfnamefont {C.}~\bibnamefont
  {Liu}}, \bibinfo {author} {\bibfnamefont {X.}~\bibnamefont {Chen}}, \ and\
  \bibinfo {author} {\bibfnamefont {L.}~\bibnamefont {Balents}},\ }\href
  {\doibase 10.1103/PhysRevB.97.245126} {\bibfield  {journal} {\bibinfo
  {journal} {Phys. Rev. B}\ }\textbf {\bibinfo {volume} {97}},\ \bibinfo
  {pages} {245126} (\bibinfo {year} {2018})}\BibitemShut {NoStop}%
\bibitem [{\citenamefont {Pal}\ and\ \citenamefont
  {Huse}(2010)}]{HusePalPRB2010}%
  \BibitemOpen
  \bibfield  {author} {\bibinfo {author} {\bibfnamefont {A.}~\bibnamefont
  {Pal}}\ and\ \bibinfo {author} {\bibfnamefont {D.~A.}\ \bibnamefont {Huse}},\
  }\href {\doibase 10.1103/PhysRevB.82.174411} {\bibfield  {journal} {\bibinfo
  {journal} {Phys. Rev. B}\ }\textbf {\bibinfo {volume} {82}},\ \bibinfo
  {pages} {174411} (\bibinfo {year} {2010})}\BibitemShut {NoStop}%
\bibitem [{\citenamefont {Serbyn}\ \emph {et~al.}(2015)\citenamefont {Serbyn},
  \citenamefont {Papi\ifmmode~\acute{c}\else \'{c}\fi{}},\ and\ \citenamefont
  {Abanin}}]{AbaninSerbynPRX2015}%
  \BibitemOpen
  \bibfield  {author} {\bibinfo {author} {\bibfnamefont {M.}~\bibnamefont
  {Serbyn}}, \bibinfo {author} {\bibfnamefont {Z.}~\bibnamefont
  {Papi\ifmmode~\acute{c}\else \'{c}\fi{}}}, \ and\ \bibinfo {author}
  {\bibfnamefont {D.~A.}\ \bibnamefont {Abanin}},\ }\href {\doibase
  10.1103/PhysRevX.5.041047} {\bibfield  {journal} {\bibinfo  {journal} {Phys.
  Rev. X}\ }\textbf {\bibinfo {volume} {5}},\ \bibinfo {pages} {041047}
  (\bibinfo {year} {2015})}\BibitemShut {NoStop}%
\bibitem [{\citenamefont {Mezard}\ \emph {et~al.}(1987)\citenamefont {Mezard},
  \citenamefont {Parisi},\ and\ \citenamefont {Virasoro}}]{Parisi1987}%
  \BibitemOpen
  \bibfield  {author} {\bibinfo {author} {\bibfnamefont {M.}~\bibnamefont
  {Mezard}}, \bibinfo {author} {\bibfnamefont {G.}~\bibnamefont {Parisi}}, \
  and\ \bibinfo {author} {\bibfnamefont {M.~A.}\ \bibnamefont {Virasoro}},\
  }\href {\doibase 10.1142/0271} {\emph {\bibinfo {title} {Spin Glass Theory
  and Beyond: An Introduction to the Replica Method and its Applications}}}\
  (\bibinfo  {publisher} {World Scientific Press},\ \bibinfo {year}
  {1987})\BibitemShut {NoStop}%
\bibitem [{\citenamefont {Abrikosov}\ \emph {et~al.}(1975)\citenamefont
  {Abrikosov}, \citenamefont {Gorkov},\ and\ \citenamefont
  {Dzyaloshinski}}]{Abrikosov1975Methods}%
  \BibitemOpen
  \bibfield  {author} {\bibinfo {author} {\bibfnamefont {A.}~\bibnamefont
  {Abrikosov}}, \bibinfo {author} {\bibfnamefont {L.}~\bibnamefont {Gorkov}}, \
  and\ \bibinfo {author} {\bibfnamefont {I.}~\bibnamefont {Dzyaloshinski}},\
  }\href {https://books.google.co.jp/books?id=E\_9NtwNY7UcC} {\emph {\bibinfo
  {title} {Methods of Quantum Field Theory in Statistical Physics}}},\ Dover
  Books on Physics Series\ (\bibinfo  {publisher} {Dover Publications},\
  \bibinfo {year} {1975})\BibitemShut {NoStop}%
\bibitem [{\citenamefont {Maldacena}\ \emph
  {et~al.}(2016{\natexlab{b}})\citenamefont {Maldacena}, \citenamefont
  {Shenker},\ and\ \citenamefont {Stanford}}]{MaldacenaDouglasJHEP2016}%
  \BibitemOpen
  \bibfield  {author} {\bibinfo {author} {\bibfnamefont {J.}~\bibnamefont
  {Maldacena}}, \bibinfo {author} {\bibfnamefont {S.~H.}\ \bibnamefont
  {Shenker}}, \ and\ \bibinfo {author} {\bibfnamefont {D.}~\bibnamefont
  {Stanford}},\ }\href {\doibase 10.1007/JHEP08(2016)106} {\bibfield  {journal}
  {\bibinfo  {journal} {Journal of High Energy Physics}\ }\textbf {\bibinfo
  {volume} {2016}},\ \bibinfo {pages} {106} (\bibinfo {year}
  {2016}{\natexlab{b}})}\BibitemShut {NoStop}%
\bibitem [{\citenamefont {Aleiner}\ \emph {et~al.}(2016)\citenamefont
  {Aleiner}, \citenamefont {Faoro},\ and\ \citenamefont
  {Ioffe}}]{AleinerAnnPhys2016}%
  \BibitemOpen
  \bibfield  {author} {\bibinfo {author} {\bibfnamefont {I.~L.}\ \bibnamefont
  {Aleiner}}, \bibinfo {author} {\bibfnamefont {L.}~\bibnamefont {Faoro}}, \
  and\ \bibinfo {author} {\bibfnamefont {L.~B.}\ \bibnamefont {Ioffe}},\ }\href
  {\doibase https://doi.org/10.1016/j.aop.2016.09.006} {\bibfield  {journal}
  {\bibinfo  {journal} {Annals of Physics}\ }\textbf {\bibinfo {volume}
  {375}},\ \bibinfo {pages} {378 } (\bibinfo {year} {2016})}\BibitemShut
  {NoStop}%
\bibitem [{\citenamefont {Jian}\ and\ \citenamefont
  {Yao}(2017)}]{YaoJianPRL2017}%
  \BibitemOpen
  \bibfield  {author} {\bibinfo {author} {\bibfnamefont {S.-K.}\ \bibnamefont
  {Jian}}\ and\ \bibinfo {author} {\bibfnamefont {H.}~\bibnamefont {Yao}},\
  }\href {\doibase 10.1103/PhysRevLett.119.206602} {\bibfield  {journal}
  {\bibinfo  {journal} {Phys. Rev. Lett.}\ }\textbf {\bibinfo {volume} {119}},\
  \bibinfo {pages} {206602} (\bibinfo {year} {2017})}\BibitemShut {NoStop}%
\bibitem [{\citenamefont {Garc\'{\i}a-Garc\'{\i}a}\ \emph
  {et~al.}(2018)\citenamefont {Garc\'{\i}a-Garc\'{\i}a}, \citenamefont
  {Loureiro}, \citenamefont {Romero-Berm\'udez},\ and\ \citenamefont
  {Tezuka}}]{TezukaGarciaPRL2018}%
  \BibitemOpen
  \bibfield  {author} {\bibinfo {author} {\bibfnamefont {A.~M.}\ \bibnamefont
  {Garc\'{\i}a-Garc\'{\i}a}}, \bibinfo {author} {\bibfnamefont
  {B.}~\bibnamefont {Loureiro}}, \bibinfo {author} {\bibfnamefont
  {A.}~\bibnamefont {Romero-Berm\'udez}}, \ and\ \bibinfo {author}
  {\bibfnamefont {M.}~\bibnamefont {Tezuka}},\ }\href {\doibase
  10.1103/PhysRevLett.120.241603} {\bibfield  {journal} {\bibinfo  {journal}
  {Phys. Rev. Lett.}\ }\textbf {\bibinfo {volume} {120}},\ \bibinfo {pages}
  {241603} (\bibinfo {year} {2018})}\BibitemShut {NoStop}%
\bibitem [{\citenamefont {Chen}\ \emph
  {et~al.}(2017{\natexlab{b}})\citenamefont {Chen}, \citenamefont {Zhou},
  \citenamefont {Huse},\ and\ \citenamefont
  {Fradkin}}]{FradkinHuseAnnPhys2017}%
  \BibitemOpen
  \bibfield  {author} {\bibinfo {author} {\bibfnamefont {X.}~\bibnamefont
  {Chen}}, \bibinfo {author} {\bibfnamefont {T.}~\bibnamefont {Zhou}}, \bibinfo
  {author} {\bibfnamefont {D.~A.}\ \bibnamefont {Huse}}, \ and\ \bibinfo
  {author} {\bibfnamefont {E.}~\bibnamefont {Fradkin}},\ }\href {\doibase
  10.1002/andp.201600332} {\bibfield  {journal} {\bibinfo  {journal} {Annalen
  der Physik}\ }\textbf {\bibinfo {volume} {529}},\ \bibinfo {pages} {1600332}
  (\bibinfo {year} {2017}{\natexlab{b}})}\BibitemShut {NoStop}%
\bibitem [{\citenamefont {Vosk}\ and\ \citenamefont
  {Altman}(2013)}]{AltmanVoskPRL2013}%
  \BibitemOpen
  \bibfield  {author} {\bibinfo {author} {\bibfnamefont {R.}~\bibnamefont
  {Vosk}}\ and\ \bibinfo {author} {\bibfnamefont {E.}~\bibnamefont {Altman}},\
  }\href {\doibase 10.1103/PhysRevLett.110.067204} {\bibfield  {journal}
  {\bibinfo  {journal} {Phys. Rev. Lett.}\ }\textbf {\bibinfo {volume} {110}},\
  \bibinfo {pages} {067204} (\bibinfo {year} {2013})}\BibitemShut {NoStop}%
\end{thebibliography}%

\end{document}